\documentclass{ar2e}

\usepackage[pdftex]{graphicx}

\begin{document}

\jname{..}
\jyear{2009}
\jvol{}
\ARinfo{1056-8700/97/0610-00}

\title{Single-Molecule Nanomagnets}

\markboth{Friedman \& Sarachik}{Single-Molecule Nanomagnets}

\author{Jonathan R. Friedman
\affiliation{Department of Physics, Amherst College, Amherst, MA 01002, USA}
Myriam P. Sarachik
\affiliation{Department of Physics, City College of New York-CUNY, New York, NY 10031, USA}}

\begin{keywords}
Single-molecule Magnets, Quantum Tunneling of Magnetization, Quantum Coherence, Berry Phase, Quantum Computation, Spintronics
\end{keywords}

\begin{abstract}
Single molecule magnets straddle the classical and quantum mechanical worlds, displaying many fascinating phenomena. They may have important technological applications in information storage and quantum computation. We review the physical properties of two prototypical molecular nanomagnets, Mn$_{12}$-acetate and Fe$_8$:
each behaves as a rigid, spin-10 object, and exhibits tunneling between up and down directions.
As temperature is lowered, the spin reversal process evolves from thermal activation to pure quantum tunneling. At low temperatures, magnetic avalanches occur in which the magnetization of an entire sample rapidly reverses. We discuss the important role that symmetry-breaking fields play in driving tunneling and in producing Berry-phase interference. Recent experimental advances indicate that quantum coherence can be maintained on time scales sufficient to allow a meaningful number of quantum computing operations to be performed. Efforts are underway to create monolayers and to address and manipulate individual molecules.
\end{abstract}

\maketitle

\newpage

\section{Introduction}

True to its name, a single-molecule magnet (SMM) is a molecule that behaves as an individual nanomagnet. Because of their small size and precise characterizability, molecular nanomagnets exhibit many fascinating quantum phenomena, such as macroscopic quantum tunneling of magnetization and Berry-phase interference. They straddle the quantum mechanical and classical worlds, residing in a middle ground that is of abiding interest to physicists. In addition, SMMs may find application in high-density magnetic storage or as qubits, the processing elements in quantum computers.

In this article, we survey some of the remarkable phenomena exhibited by SMMs and discuss progress towards future applications. In Section 2, we review the basic structure and properties of SMMs with reference to the two prototypical molecules, Mn$_{12}$-acetate and Fe$_8$, and provide a brief history. In Section 3, we discuss the reversal of the magnetic moment by quantum tunneling; the crossover from classical spin reversal to pure quantum tunneling; the symmetry-breaking fields that drive tunneling; and the abrupt reversal of the magnetic moment of an entire crystalline sample in the form of a magnetic avalanche. The experimental observation of geometric-phase (Berry-phase) interference is described in Section 4. In Section 5, we discuss recent developments that show that quantum coherence can be maintained in SMMs on time scales sufficient to allow a significant number of qubit operations to be performed. These exciting results make SMMs serious contenders for use in quantum information technologies. In Section 6, we discuss recent experimental efforts to create single layers of SMMs on surfaces and to measure transport through individual molecules.

\section{Background}

\begin{figure}[tb]
\centering
\includegraphics[width=1\linewidth]{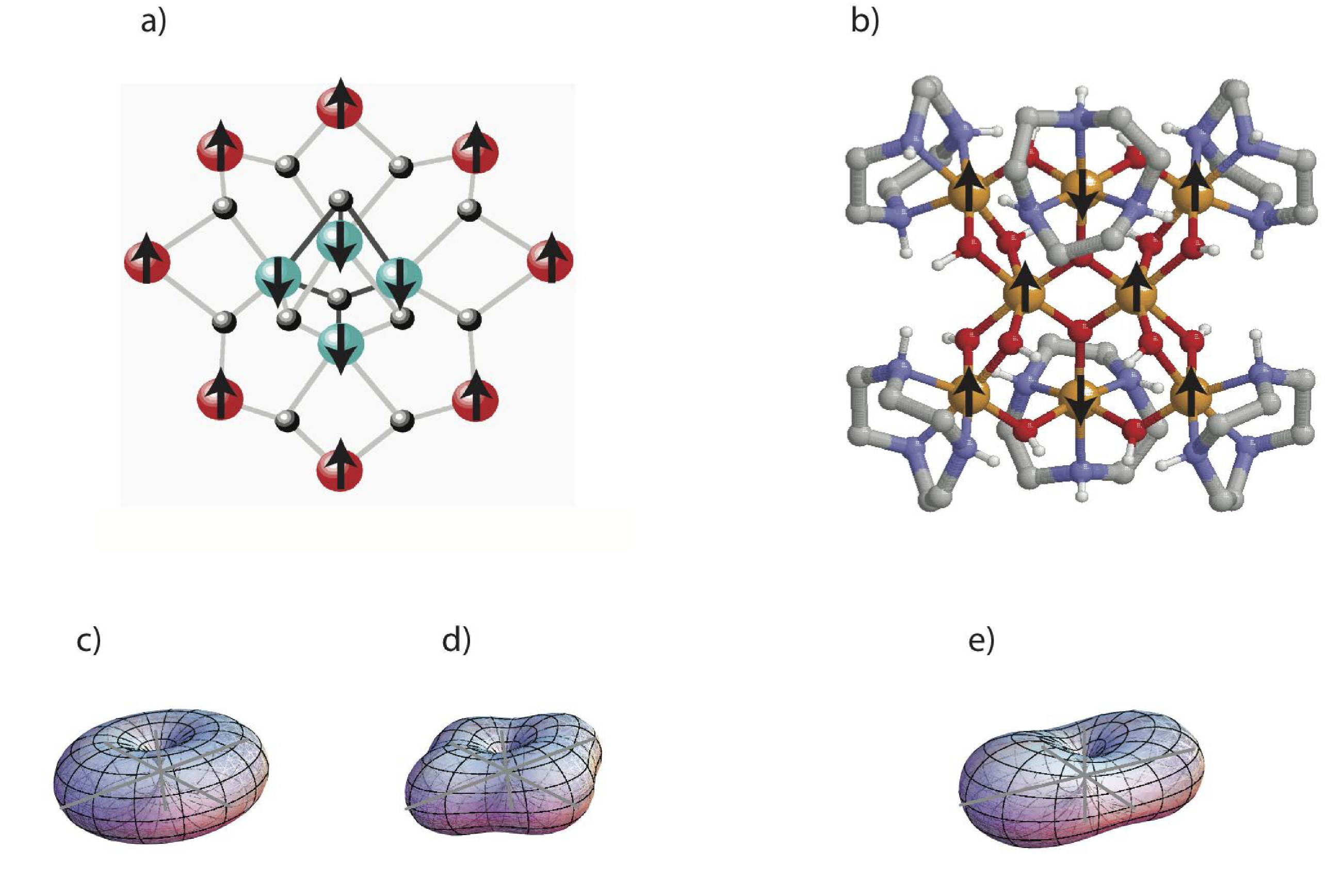}
\caption{(a) Chemical structure of the core of the Mn$_{12}$ molecule. The four inner spin-down Mn$^{3+}$ ions (green) each have spin $S=3/2$; the eight outer (red) spin-up Mn$^{4+}$ ions each have spin $S = 2$, yielding a net spin $S = 10$ for the magnetic cluster; black dots are O bridges and arrows denote spin. Acetate ligands and water molecules have been removed for clarity. (b) Structure of Fe$_8$. Brown atoms represent Fe, with arrows denoting spin; red circles are O, mauve circles are N, and gray circles are C. Br, H, and ligands are not shown. (c) Spherical polar plots of energy as a function of orientation for a classical spin with uniaxial anisotropy; (d) same as (c) with additional fourth-order transverse anisotropy. (e) Spherical polar plot of energy for a spin with biaxial anisotropy.}
\label{fig1}
\end{figure}

The spin of a SMM ranges from a few to many times that of an electron; the corresponding magnetization of the individual magnets is minuscule. The molecules readily crystallize so that a typical sample contains $
\sim10^{15}$ or more identical magnetic clusters in (nearly) identical crystalline environments. At the same time, the SMMs are relatively far apart so that the magnetic exchange between them is small and they interact only very weakly with each other. To a very good approximation, a crystalline sample thus behaves at low temperatures as an ensemble of well-characterized, identical, non-interacting nanoscale magnets. Although the symmetry, the magnitude of spin anisotropy, as well as the hyperfine fields, dipolar interactions and other properties, vary substantially from one SMM to another, most exhibit the same overall behavior. The central features can be understood with reference to the prototypical SMMs Mn$_{12}$-ac and Fe$_8$ shown in Figure~\ref{fig1}.

First synthesized by Lis in 1980 \cite{lis}, Mn$_{12}$O$_{12}$(CH$_3$COO)$_{16}$(H$_2$O)$_4$ (referred to hereafter as Mn$_{12}$-ac) received little attention until its unusually large molecular magnetic moment \cite{spin-10} and magnetic bistability \cite{bistability} were recognized. Early measurements established a number of important features: a large $S=10$ spin, rigid at low temperatures; a large negative magnetocrystalline anisotropy with a barrier $U \sim 70$ K \cite{spin-10,anisotropy}, resulting in a characteristic relaxation time $\tau$ that obeys an Arrhenius law and magnetic hysteresis below a ``blocking temperature" $T_B \sim 3$ K \cite{bistability,hysteresis,87,50,Paulsen,13,14}.

\begin{figure}[tb]
\centering
\includegraphics[width=1\linewidth]{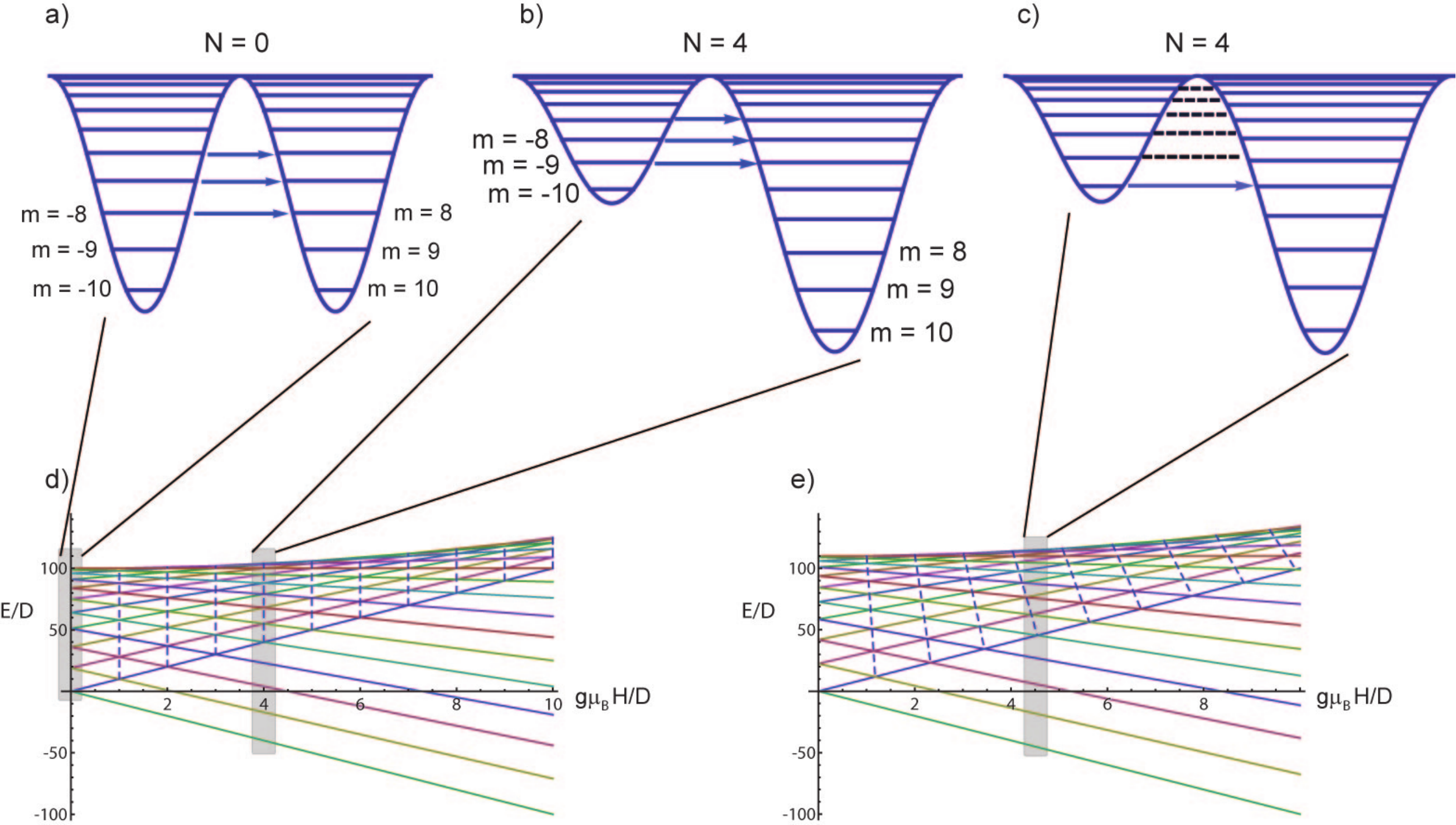}
\caption{(a) Double-well potential in the absence of magnetic field showing spin-up and spin-down levels separated by the anisotropy barrier. Different spin projection states $|m>$ are indicated for $S=10$. The arrows denote quantum tunneling. (b) Double-well potential for the N=4 step when a magnetic field is applied along the easy axis. (c) Double-well potential with a small fourth-order term in the Hamiltonian, $\propto S_z^4$; due to the presence of this term, different pairs of energy levels coincide at different magnetic fields. (d) Energy of spin projection states $|m>$ versus applied magnetic field derived from the first two terms of the Hamiltonian, Eq.~\ref{HamTam}. The vertical dashed lines indicate that all level pairs cross simultaneously in this simple approximation. (e) Energy levels as a function of magnetic field when a fourth-order term is included. As denoted by the dashed lines, different level pairs are in resonance for different magnetic fields within a given step $N$.}
\label{fig2}
\end{figure}

As shown in Figure~\ref{fig1}(a), the magnetic core of Mn$_{12}$-ac has four Mn$^{4+}$ (S = 3/2) ions in a central tetrahedron surrounded by eight Mn$^{3+}$ (S = 2) ions. The ions are coupled by superexchange through oxygen bridges with the net result that the four inner and eight outer ions point in opposite directions, yielding a total spin $S=10$. The magnetic core is surrounded by acetate ligands, which serve to isolate each core from its neighbors and the molecules crystallize into a body-centered tetragonal lattice. While there are very weak exchange interactions between molecules, the exchange between ions within the magnetic core is very strong, resulting in a rigid spin$-10$ object that has no internal degrees of freedom at low temperatures. As illustrated by Figure~\ref{fig2}(a), the spin's energy can be modeled as a double-well potential, where one well corresponds to the spin pointing ``up'' and the other to the spin pointing ``down''. A strong uniaxial anisotropy barrier of the order of 70 K yields doubly degenerate ground states in zero field. The spin has a set of levels (shown in the figure) corresponding to different projections, $m = 10, 9,\ldots, -9, -10$, of the total spin along the easy axis of the molecule (corresponding to the $c$-axis of the crystal).

The magnetic cluster [(tacn)$_6$Fe$_8$O$_2$(OH)$_{12}$]Br$_8$ (referred to as Fe$_8$) is shown in Figure~\ref{fig1}(b); here (tacn) is the organic ligand $1$,$4$,$7$-triazacyclononane. Like Mn$_{12}$, Fe$_8$ has a spin ground state $S = 10$, which arises from competing antiferromagnetic interactions between the eight $S = 5/2$ Fe spins. Modeled by a double-well potential like that of Mn$_{12}$ (see Figure~\ref{fig2}), the spin dynamics are quite similar. However, as shown by the schematic diagrams of Figure~\ref{fig1}(c-e), while Mn$_{12}$-ac has an easy axis and an essentially isotropic hard plane, Fe$_8$ has three inequivalent axes. The fact that Fe$_8$ is biaxial rather than uniaxial has interesting consequences that are presented in Section 4.

\section{Macroscopic Quantum Tunneling of Magnetization}

A number of enigmatic features emerged in the mid-1990's that suggested the possibility of spin reversal by quantum mechanical tunneling in Mn$_{12}$-ac. In 1995, Barbara {\em et al.}~\cite{87} observed an increase in the magnetic relaxation time $\tau (H)$ when a longitudinal magnetic field was increased from zero to $\sim 0.2$ T, above which the relaxation decreased. This is counterintuitive and puzzling, since the application of a field lowers the barrier for spin reversal so that the relaxation time is expected to decrease monotonically as a function of field. Barbara {\em et al.}~\cite{87} suggested that the faster relaxation at zero field could be due to ``the coincidence of the level schemes of the two wells.'' At about the same time, Novak and Sessoli \cite{50} reported relaxation minima at $H = 0$ and $0.3$ T; they speculated that this might be due to thermally assisted tunneling between excited states in a double-well potential. Paulsen and Park \cite{Paulsen} reported magnetic avalanches (rapid, complete magnetization reversals) that occurred most often at a specific field of $\sim 1$ T. These experiments all found enigmatic behavior at particular values of the magnetic field, suggesting the possibility of quantum tunneling.

\begin{figure}[tb]
\centering
\includegraphics[width=1\linewidth]{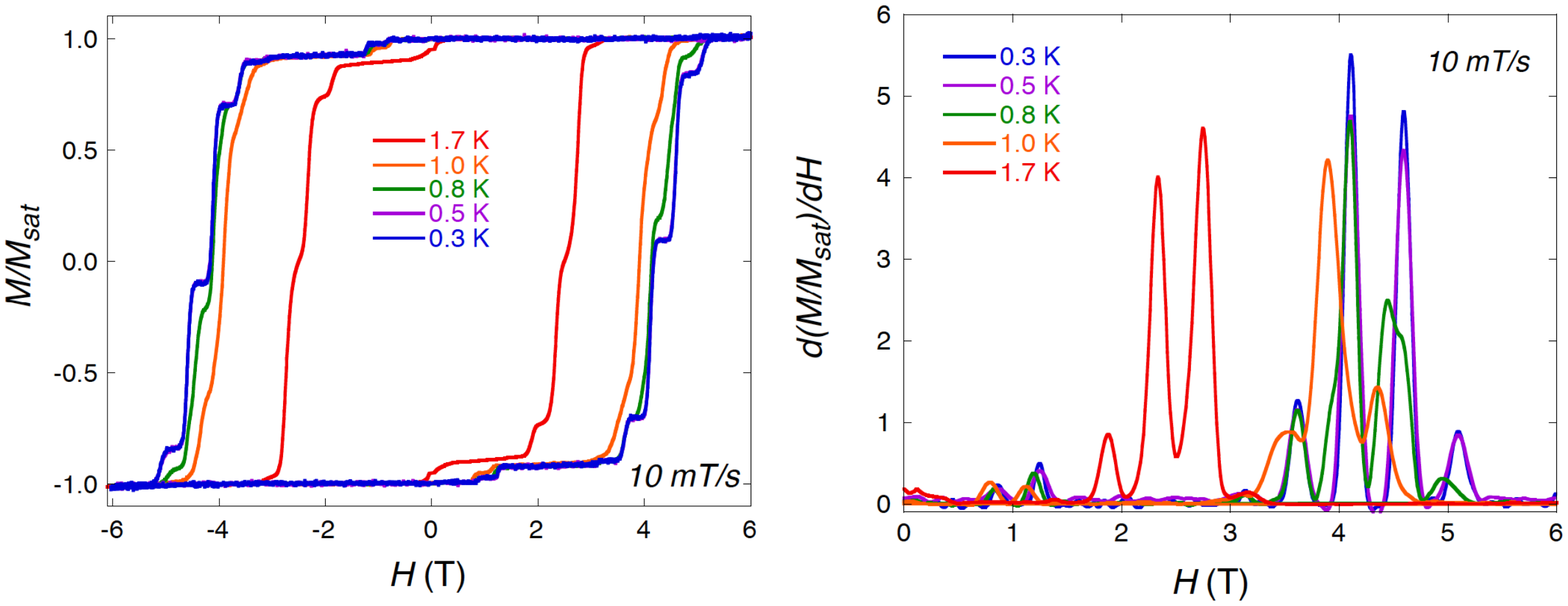}
\caption{(a) Magnetization of a Mn$_{12}$-ac crystal normalized by its saturation value as a function of magnetic field applied along the uniaxial $c$-axis direction at different temperatures below the blocking temperature; the magnetic field was swept at 10 mT/s. (b) The derivative, $dM/dH_z$ of the data in part (a) as a function of magnetic field.}
\label{fig3}
\end{figure}

The observation by Friedman {\em et al.}~\cite{friedman} of macroscopic quantum tunneling of the magnetization in Mn$_{12}$-ac in 1996 and its confirmation shortly thereafter by Hern\'{a}ndez {\em et al.}~\cite{89} and Thomas {\em et al.}~\cite{Thomas} is widely recognized as a major breakthrough in spin physics \cite{nature96}. A series of steps were discovered in the hysteresis loops in Mn$_{12}$-ac below the blocking temperature of $\sim 3$ K; typical curves are shown in Figure~\ref{fig3}. Figure~\ref{fig3}(a) shows the magnetization $M$ as a function of magnetic field $H_z$ applied along the easy axis; the derivative, $dM/dH_z$, which reflects the magnetic relaxation rate, is plotted as a function of $H_z$ in Figure~\ref{fig3}(b). As shown in Figure~\ref{fig2}, $H_z$ tilts the potential, causing levels in the right (left) well to move down (up). Figure~\ref{fig2}(d) shows the field dependence of the spin's energy levels. Levels in opposite wells align at certain values of magnetic field (dashed lines in Figure~\ref{fig2}(d)), allowing the spin to reverse by resonant tunneling. The steps observed in the hysteresis loops at nearly equal intervals of magnetic field are due to enhanced relaxation of the magnetization at the resonant fields when levels on opposite sides of the anisotropy barrier coincide in energy. This magnetization tunneling phenomenon has now been seen in hundreds of SMMs as well as in some high-spin rare-earth ions \cite{551,756a}.

SMMs generally owe their simplicity to the fact that they behave as single, rigid spins at sufficiently low temperatures. The effective spin Hamiltonian can be written as:
\begin{equation}
{\cal H} \approx - DS_z^2 - g_z\mu_B S_z H_z - AS_z^4 + {\cal H^\prime} \label{HamTam},
\end{equation}

\noindent where the first term gives rise to the anisotropy barrier; the second term is the Zeeman energy that splits the spin-up and spin-down states in a magnetic field, thereby lifting the degeneracy of the two potential wells; the third is the next-highest-order term in longitudinal anisotropy; and the last term, $\cal H^\prime$, contains all symmetry-breaking operators that do not commute with $S_z$. Note that in the absence of $ {\cal H^\prime}$, $S_z$ is a conserved quantity and no tunneling would be allowed. For Mn$_{12}$-ac, $D=0.548$K, $g_z = 1.94$, and $A=1.173 \times 10^{-3}$ K.

The main features of the behavior of Mn$_{12}$-ac, a particularly simple and highly symmetric SMM, can be understood by considering only the first and second terms in Eq.~\ref{HamTam}; the first creates the double-well potential and the second provides the tilt of the potential in a magnetic field. The steps observed in Figure~\ref{fig3} occur at resonant fields $H_N = ND/g_z\mu_B$ at which levels in opposite wells (Figure~\ref{fig2}) align. The fields at which the steps occur are consistent with the independently measured values of $D$ and $g_z$ \cite{ESR}; the steps are labeled sequentially by integers starting with $N=0$ for $H=0$. At temperatures below $0.5$ K, tunneling proceeds predominantly from the ground state of the metastable well: as shown in Figure~\ref{fig3}, the hysteresis loops are essentially identical for $T=0.3$ K and $T=0.5$ K. As the temperature increases the hysteresis loops become narrower and magnetic relaxation occurs at lower magnetic fields corresponding to smaller values of $N$: the spins are activated to higher levels from which they can more easily tunnel across the barrier. This thermally assisted tunneling process can be described using a master-equation approach in which transitions between levels occur by tunneling or by the absorption or emission of phonons \cite{24,221,113,114,168,662,656,garaninlong}. Above the blocking temperature $T_B$ (a phenomenological parameter that depends on the time scale of the measurement), sufficient thermal energy is available for the magnetization to quickly achieve equilibrium and no hysteresis is observed. This two-term Hamiltonian provides a good description of many other SMMs as well.

Although they are small, additional terms in the Hamiltonian are responsible for important details that have provided many new insights. We will introduce these additional terms one by one, and discuss their consequences.

For the two-term Hamiltonian with a simple quadratic anisotropy, $-DS_z^2$, the level crossings corresponding to each resonance occur pairwise at the same value of magnetic field (Figure \ref{fig2}(d)). That is, every level in the left well simultaneously crosses a level in the right well at one value of field. A $S_z^4$ term, determined by electron spin resonance (ESR) \cite{ESR} and inelastic neutron scattering \cite{neutrons}, removes this coincidence and introduces detailed structure in each step.

The magnetic field corresponding to spin tunneling from level $m'$ of the metastable well to $m$ in the stable well can be easily calculated for the three-term Hamiltonian ${\cal H} = - DS_z^2 - g_z\mu_B S_z H_z - AS_z^4$, yielding:

\begin{equation}
H_{m',m} = \frac{D(m'+m)}{g_z \mu_B}\left[1+\frac{A}{D}(m'^2+m^2)\right],
\end{equation}

\noindent where $(m'+m)$ is the step number $N$. While the first term in the brackets gives resonant magnetic fields that are integer multiples of $D/g_z\mu_B$ independently of the pair ($m',m$), as illustrated in Figure~\ref{fig2}(b and d), the small correction $A/D = 2.1 \times 10^{-3}$ causes different pairs of levels to cross at different values of the magnetic field within each step, as shown in Figure~\ref{fig2}(c and e). This effect becomes increasingly important for larger step numbers (high magnetic field), as shown by the dashed lines in Figure~\ref{fig2}(e).

\begin{figure}[tb]
\centering
\includegraphics[width=1\linewidth]{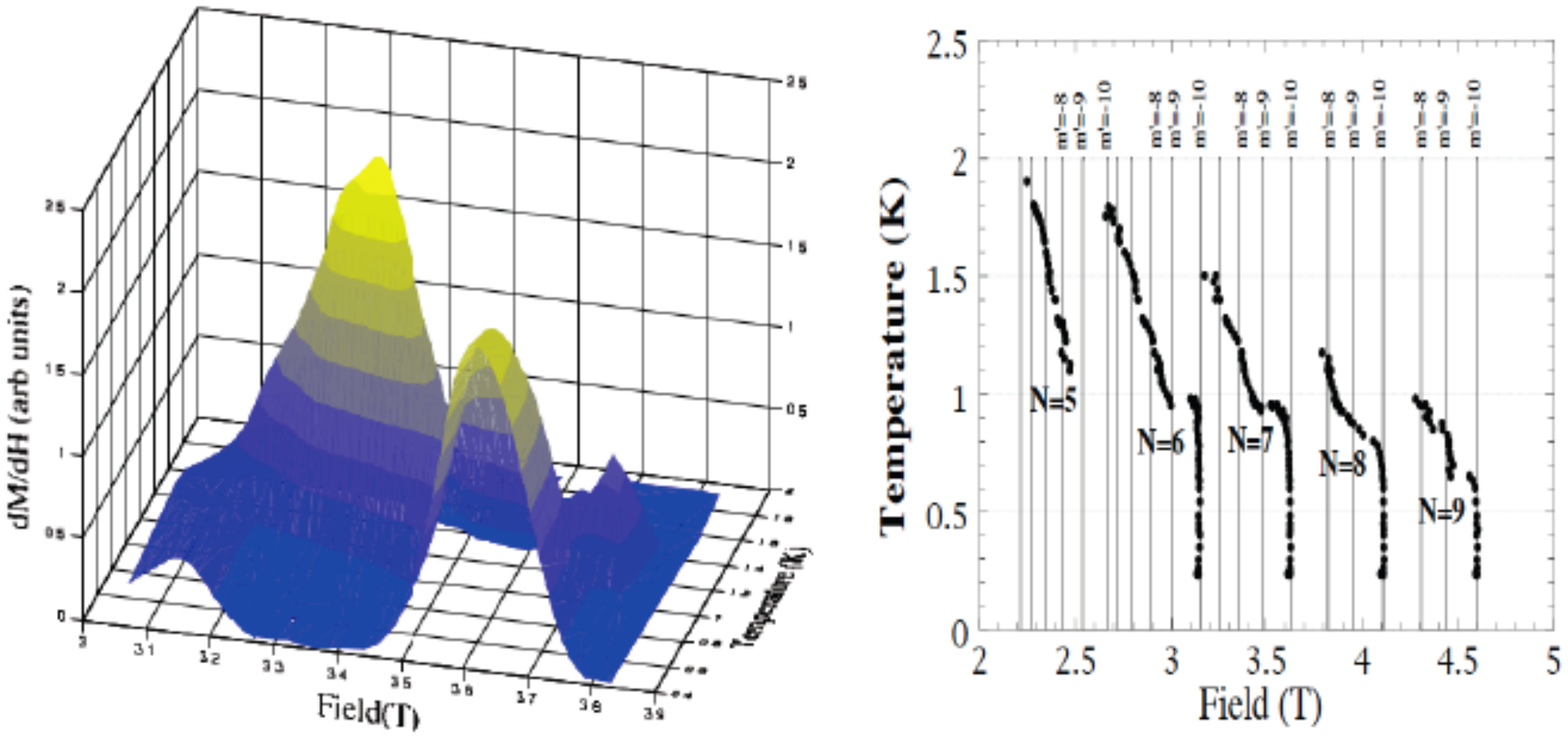}
\caption{(a) $dM/dH_z$ as a function of magnetic field and temperature for tunneling step $N=7$. As shown in Figures 2c and 2e, different level pairs $(m',m)$ within a given step are in resonance at different magnetic fields, so that enhanced relaxation due to quantum tunneling shifts to different pairs as the field and temperature vary. Note the abrupt transfer of weight from thermally-assisted tunneling (left-hand peak) to pure quantum tunneling from the lowest state of the metastable potential well (right-hand peak). (b) The abrupt transfer of weight shown for several steps, as indicated.}
\label{fig4}
\end{figure}

This feature provides an interesting form of spectroscopy that allows a determination of which energy levels are responsible for tunneling. As the temperature is reduced, the relaxation evolves from thermally assisted tunneling (i.e., thermal activation to a higher level from which tunneling proceeds) to tunneling from the lowest $m=-10$ level of the metastable well. The crossover between these regimes is shown in Figure~\ref{fig4}. It is abrupt rather than gradual, suggesting a (discontinuous) first-order rather than a (continuous) second-order transition \cite{abrupt1,abrupt2,abrupt3}. We note that a true first-order, discontinuous transition occurs only in the limit of infinite spin; the transition in Mn$_{12}$-ac, where the spin $S=10$ is large but finite, is thus abrupt rather than discontinuous.

In order for tunneling to occur, the Hamiltonian must include terms that do not commute with $S_z$, which we have collectively labeled ${\cal H^\prime}$:
\begin{equation}
{\cal H^\prime} \label{HamTam2} = E(S_x^2 - S_y^2) - g\mu_BH_x S_x + (C/2)(S_+^4 + S_-^4) + \ldots
\end{equation}
The first term on the right-hand side is a second-order transverse anisotropy that is present in many low-symmetry SMMs; the second term includes hyperfine, dipolar, and possibly other internal transverse fields as well as an externally applied transverse field; the third term is the fourth-order transverse anisotropy.

The source of tunneling in Mn$_{12}$-ac was the subject of intense debate and investigation for a number of years. In a perfect crystal, the lowest transverse anisotropy term allowed by the tetragonal symmetry of Mn$_{12}$-ac is $(C/2)(S_+^4 + S_-^4)$. This imposes a selection rule in which $m$ can change only by integer multiples of 4, $\Delta m = 4 i, i=0,1,\ldots$, allowing only every fourth step for ground-state tunneling. For thermally assisted tunneling, this selection rule prohibits every other step \cite{Hernandez}. By contrast, all steps are observed with no clear differences in amplitude between them (see Figure~\ref{fig3}). Dipolar fields and hyperfine interactions, which would allow all steps on an equal footing, are known to be too weak to cause the rapid tunneling rates observed experimentally \cite{Hernandez}.

Through a series of theoretical \cite{chudnovskygaranin,park04} and experimental \cite{mertesdistribution,kentdistribution1,kentdistribution2,hilldistribution1,hilldistribution2} steps, the source of tunneling has been traced to isomer disorder \cite{cornia} in Mn$_{12}$-ac. Specifically, variation in the hydrogen bonding of Mn$_{12}$ molecules with neighboring acetic acid molecules leads to a distribution of quadratic (rhombic) transverse anisotropy. This introduces a locally varying second-order anisotropy superposed on the global tetragonal symmetry of the crystal and induces tunneling through the first term in ${\cal H^\prime}$, Eq.~\ref{HamTam2}. The symmetry of such a term permits tunneling at all even-numbered steps. In addition, the isomer disorder produces a distribution of tilts (within $\approx 1.7^\circ$) of the molecular easy axes with respect to the global uniaxial direction, the crystal's $c$ axis. When a field is applied along this axis, the tilt distribution gives rise to a distribution of transverse fields that drives tunneling by virtue of the term linear in $H_x$ (in Eq.~\ref{HamTam2}) and allows all steps, both even and odd.

\begin{figure}[tb]
\centering
\includegraphics[width=1\linewidth]{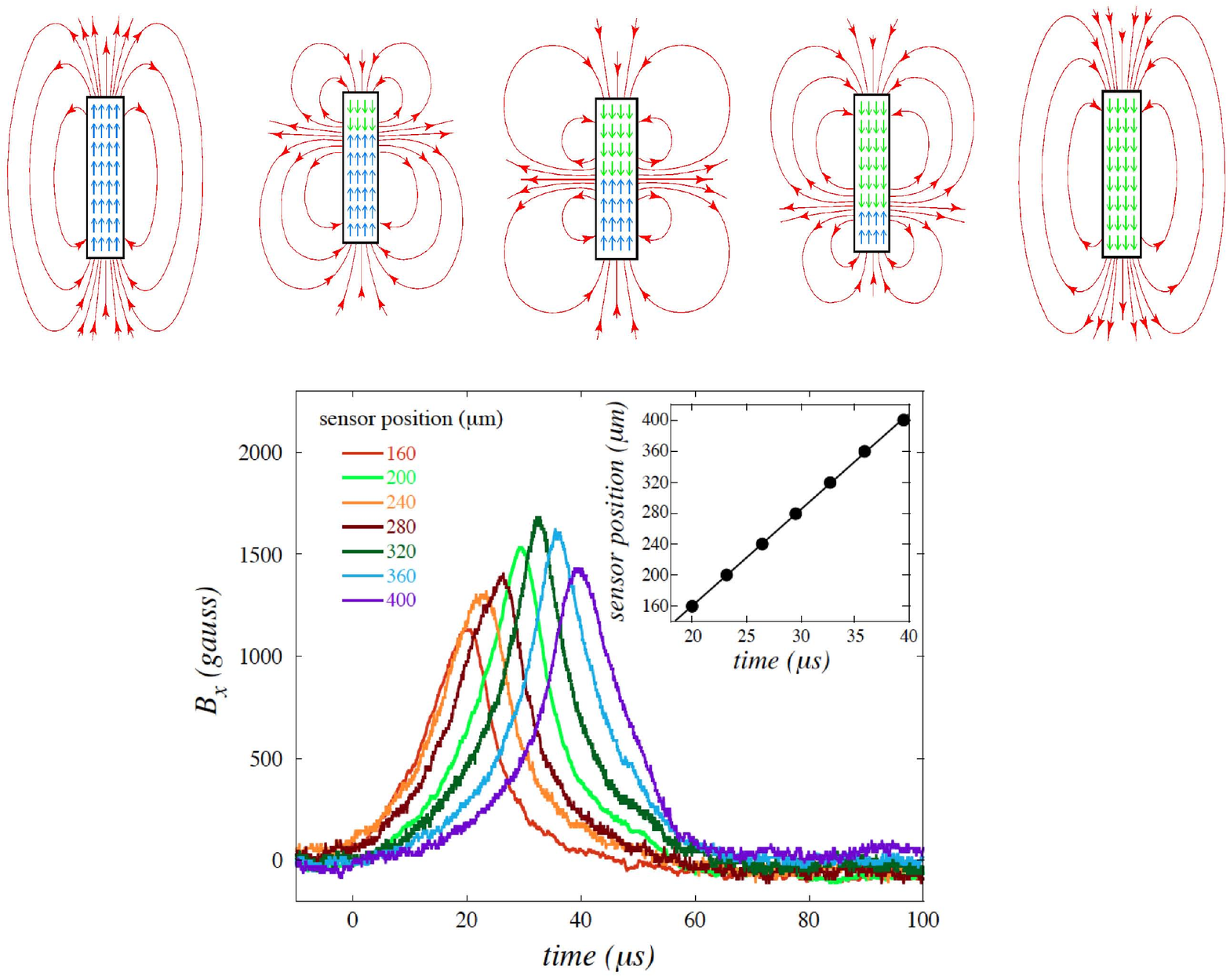}
\caption{Upper panel: Schematic diagram of magnetic field lines as spins reverse direction during a magnetic avalanche traveling from top to bottom of a Mn$_{12}$ crystal. Lower panel: The local magnetization measured as a function of time by an array of micron-sized Hall sensors placed along the surface of the sample. Each peak corresponds to the ``bunching" of the magnetic field lines as the deflagration front travels past a given Hall sensor. The propagation speed for this avalanche is $10$ m/s, approximately two orders of magnitude below the speed of sound.}
\label{fig5}
\end{figure}

We end this section with a brief description of the process of spin reversal by magnetic avalanches in molecular magnets. As first reported by Paulsen and Park \cite{Paulsen} in Mn$_{12}$-ac, crystals of molecular magnets often exhibit an abrupt and complete reversal of the magnetization from one direction to the other. Poorly understood until recently, these avalanches were attributed to a thermal runaway in which heat is released that further accelerates the magnetic relaxation. In addition to releasing thermal energy, molecular crystals emit bursts of radiation during magnetic avalanches \cite{tejadaradiation1,tejadaradiation2,keren}. Once considered events to be avoided, as they interfere with a detailed study of the stepwise process of magnetization reversal, magnetic avalanches became the focus of attention and renewed interest stimulated by the theoretical suggestion that the radiation emitted during an avalanche is in the form of coherent (Dicke) superradiance \cite{superradiance}. Although the issue of coherence of the radiation has yet to be resolved, recent studies have clarified the nature of the avalanche process itself. In particular, local time-resolved measurements using micron-sized Hall sensors have shown that a magnetic avalanche spreads as a narrow interface that propagates through the crystal at a constant velocity that is roughly two orders of magnitude smaller than the speed of sound \cite{suzuki}. This process, illustrated schematically in Figure~\ref{fig5}, is closely analogous to chemical combustion - the propagation of a flame front through a flammable chemical substance, referred to as chemical deflagration.

\begin{figure}[tb]
\centering
\includegraphics[width=1\linewidth]{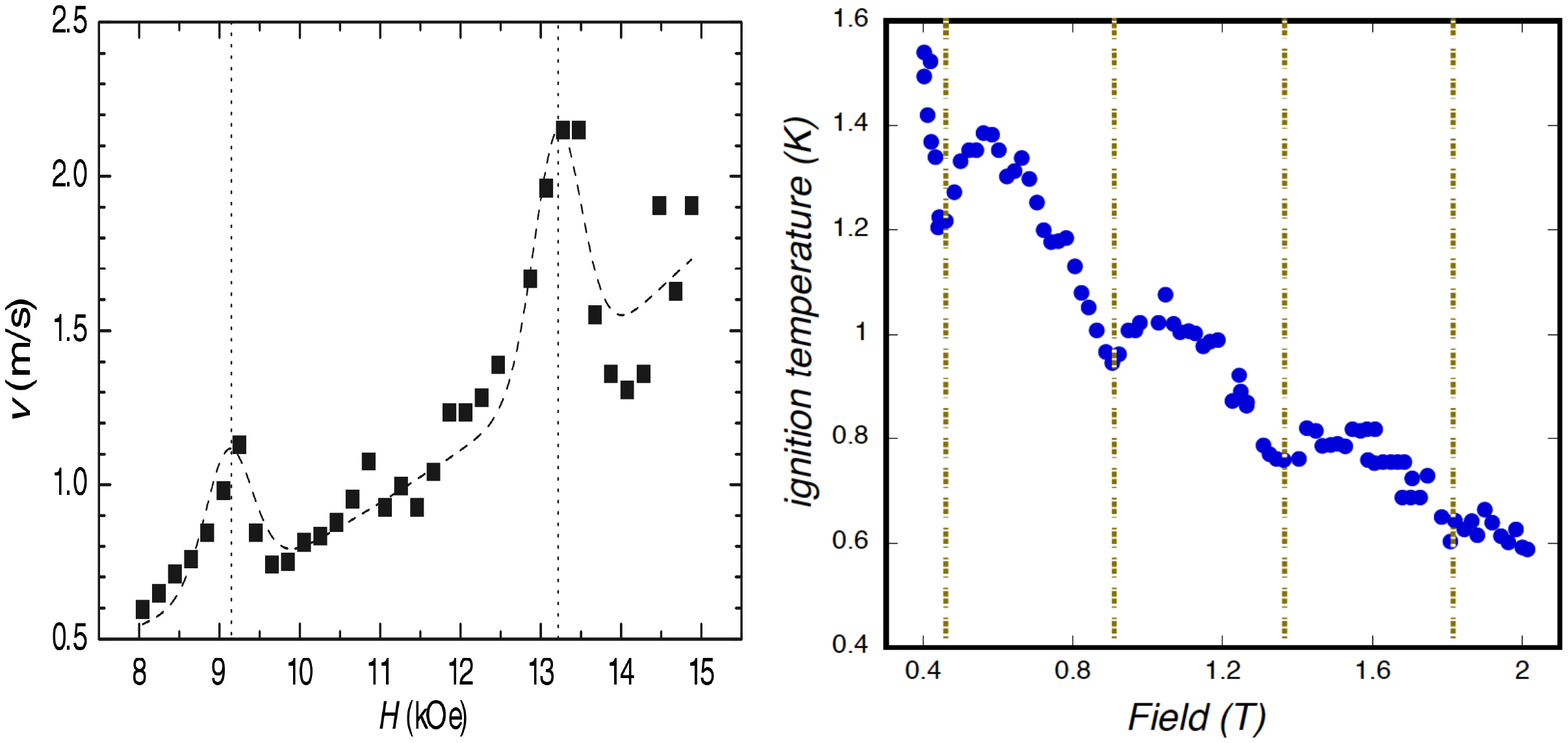}
\caption{(a) The speed of propagation of the magnetic avalanche deflagration front is plotted as a function of the field at which the avalanche is triggered; note the enhancement of propagation velocity at magnetic fields corresponding to quantum tunneling (denoted by vertical dotted lines). From Ref.~\cite{tejadapeaks1}; reproduced with permission. (b) The temperature above which an avalanche is ignited displays clear minima at the same resonant fields (vertical dashed lines).}
\label{fig6}
\end{figure}

Interestingly, there is clear evidence of the quantum-mechanical nature of the spin-reversal process during an avalanche. Magnetic avalanches have been studied in detail by time-resolved measurements of the local magnetization \cite{seandips} and by measurements of bulk magnetization during avalanches ignited by surface acoustic waves \cite{tejadapeaks1,tejadapeaks2}, as well as theoretically by Garanin and Chudnovsky \cite{avalanchetheory}. As shown in Figure \ref{fig6}(a), the avalanche speed is enhanced at the resonant values of magnetic field at which tunneling occurs \cite{seandips,tejadapeaks1,tejadapeaks2}; Figure \ref{fig6}(b) shows that at the same resonant fields there are pronounced dips \cite{seandips} in the temperature required to ignite an avalanche.

\section{Berry Phase in Molecular Magnets}

While we have so far restricted our discussion to the Mn$_{12}$-ac molecule, many other SMMs exhibit similar behavior. As discussed above, the Fe$_8$ molecule resembles Mn$_{12}$: it has a spin of 10 \cite{193}, a substantial anisotropy barrier ($\approx $22 K) and shows resonant tunneling steps in the hysteresis loops \cite{91}.

\begin{figure}[tb]
\centering
\includegraphics[width=0.5\linewidth]{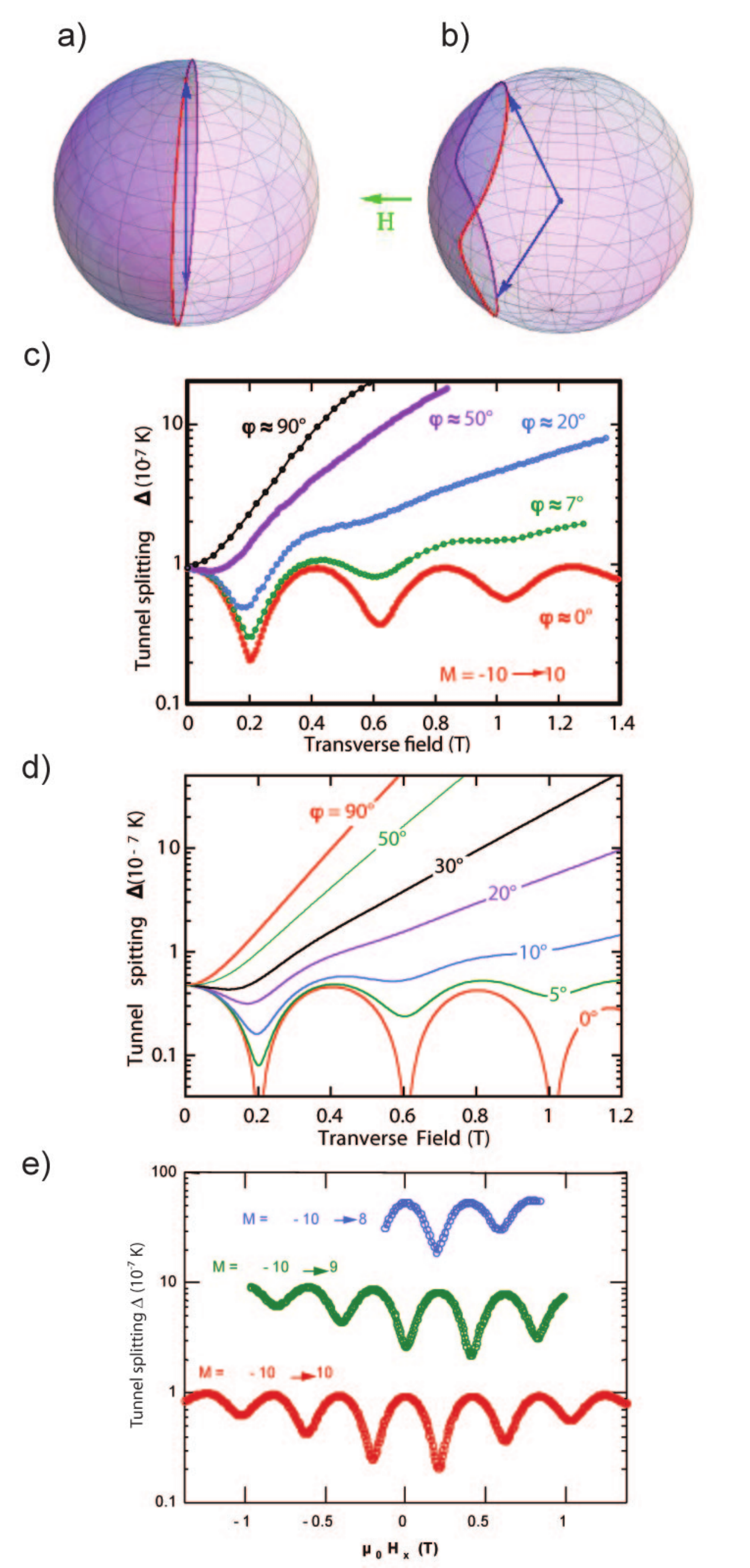}
\caption{(a) Bloch sphere for spin tunneling in zero field. The blue arrows indicate the ground-state directions of the spin (corresponding to minima of the potential wells in Figure~\ref{fig2}(a)). The two least-action (instanton) tunneling paths are indicated by the red and purple curves. The geometric phase that produces the interference is proportional to the solid angle subtended by the surface spanning the two paths (shaded hemisphere). (b) When a field {\em H} is applied along the hard $x$ axis, the two ground-state orientations are tilted toward that axis, altering the tunneling paths (red and purple), the solid angle between them (shaded region), and the interference between the paths. (c) Measured tunnel splittings for Fe$_8$ when the field is applied along the hard axis ($\phi=0$) and in several other directions in the hard plane, as labeled. (d) Calculated tunnel splitting for the transverse field applied along various directions in the hard plane (including both second- and fourth-order transverse anisotropies). (e) Measured tunnel splittings for Fe$_8$ for a transverse field applied along the hard axis for three different tunnel resonances, as indicated. Panels c-e from \cite{162}; reproduced with permission.}
\label{fig7}
\end{figure}

As shown schematically in Figure~\ref{fig1}(e), Fe$_8$ has three inequivalent directions, providing a hard $x$ axis and a ``medium" $y$ axis within the hard plane. In this case, it is essential to retain the first term in Eq.~\ref{HamTam2}, namely, $E(S_x^2 - S_y^2) $. The presence of this term indicates that in zero magnetic field the spin has two preferred tunneling paths that ``pass through" the $y$ and $-y$ directions, respectively, as illustrated by the red and purple curves in Figure~\ref{fig7}(a). This leads to a remarkable interference effect.

Geometric (or Berry) phase is a fascinating phenomenon in both classical and quantum physics in which a system adiabatically following a closed path in some parameter space acquires a non-trivial phase change. A familiar example of a geometric phase is the Aharanov-Bohm effect in which a charged particle whose path encircles a region of magnetic flux acquires a phase proportional to the flux.

When a spin's orientation traverses a closed path, it acquires a geometric phase proportional to the solid angle enclosed by that path: $\int (1-cos \theta)d\phi$, where $\theta$ and $\phi$ are the polar and azimuthal angles describing the position of the spin's vector on the unit sphere. A biaxial spin, like Fe$_8$, has two least-action tunneling paths for spin reversal. Each path acquires a different geometric phase and these paths will therefore interfere.

In 1993, well before the discovery of tunneling in SMMs, Garg \cite{166,63} predicted that a magnetic field could be used to modulate the geometric-phase interference. While a field oriented along the hard ($x$) axis preserves the symmetry between the two tunneling paths -- they both maintain the same amplitude -- it changes the geometric phase difference between the paths, altering the interference.

The tunnel splitting is modulated by the factor cos$(S\Omega$), where $\Omega$ is the solid angle circumscribed by the two paths, as illustrated by the shaded regions in Figure~\ref{fig7}(a) and ~\ref{fig7}(B). As the magnitude of the field is increased, $\Omega$ decreases and whenever $S\Omega = (2n+1) \times \pi/2$ for integer $n$, the interference is completely destructive, causing the tunnel splitting to vanish. The predicted field interval between zeros is

\begin{equation}
{\Delta H} = \frac{2}{g\mu_B} \sqrt{2E(E+D)}.
\end{equation}

This interference effect was first discovered experimentally in Fe$_8$ in 1999 by Wernsdorfer and Sessoli \cite{162}. They used a Landau-Zener tunneling method, in which a longitudinal field sweeps the system rapidly through a tunneling resonance, to determine the tunnel splitting for that resonance. The results, plotted in Figure~\ref{fig7}(c), show clear oscillations for the $N$ = 0 resonance when the field is applied along the hard axis ($\phi=0$). The tunnel splitting drops by nearly an order of magnitude at regular field intervals of $0.41$ T. The effect becomes less pronounced as the field in the hard plane is tilted away from the hard axis. Qualitative agreement with theory is obtained, and there is quantitative agreement when one includes an additional, fourth-order transverse anisotropy term in the spin Hamiltonian for Fe$_8$ (the third term in Eq.~\ref{HamTam2}), as shown in Figure~\ref{fig7}(d). Garg's original calculation considered the case of a purely transverse field (i.e. $N$=0). As Figure~\ref{fig7}(e) illustrates, the interference was also observed at other resonances ($N$ = 1 and 2). This unexpected finding prompted much theoretical work to understand the observations \cite{607, 593, 601}.

Another interesting feature of the data is illustrated in Figure~\ref{fig7}(e), which shows that the odd resonance ($N = 1$) is out of phase with the even resonances ($N = 0, 2$). That is, when the tunnel splitting of an even resonance has a maximum, the tunnel splitting of the odd resonance has a minimum and vice-versa. This so-called parity effect is a manifestation of a selection rule imposed by the second-order transverse anisotropy for Fe$_8$, namely, that odd resonances are forbidden for an integer spin. The fact that the tunneling does not go to zero exactly is likely due to dipolar interactions between spins that produce local transverse fields, an effect that has been demonstrated experimentally \cite{181}.

Since the work of Wernsdorfer and Sessoli, the geometric-phase effect has been observed in other SMMs. Most of these have an effective Hamiltonian like that of Fe$_8$ with a single hard-axis direction \cite{595, 434}. Geometric-phase interference has recently been observed in a Mn$_{12}$ variant (Mn$_{12}$-tBuAc) that has a fourth-order transverse anisotropy of the form $\frac{C}{2}(S_+^4 + S_-^4)$, in which the system has four hard-axis directions ($\pm x$ and $\pm y$ -- see Figure\ref{fig1}(d)) \cite{747}, as had been predicted theoretically \cite{414, 216}. Another interference effect has been reported in systems that behave as exchange-coupled dimers of SMMs \cite{697, 681}, where it is the effective exchange interaction that is modulated by an applied field, but a full theoretical understanding of the effect has not yet been found. Geometric-phase interference has recently been observed in an antiferromagnetic SMM \cite{749a}, a manifestation of interference between N\'{e}el-vector tunneling paths \cite{750}. There has also been recent theoretical work that predicts that uniaxial stress applied the along the hard axis of a four-fold symmetric SMM, like Mn$_{12}$-tBuAc, will produce a geometric-phase effect in the absence of a magnetic field \cite{748}.

\section{Quantum Coherence; Quantum Computation}

Of the many potential applications for SMMs, one of the most interesting is the possibility that they could be used as qubits, the processing elements of quantum computers.
Quantum computers exploit uniquely quantum properties like superpositions of states and entanglement. They can, in principle, solve certain problems, like factoring large numbers, more efficiently than could be done with any known algorithm for a classical computer \cite{305}. The basic processing elements of quantum computers are qubits, which like classical bits, can be put into logical $|0>$ and $|1>$ states. Unlike classical bits, however, they can also be put into superposition states, i.e. a$|0>$ + b$|1>$. Qubits can also be entangled with one another where the state of an individual qubit is ill-defined.

Many physical systems have been proposed as possible qubits, from microscopic systems like trapped ions \cite{739} and nuclear spins \cite{305} to macroscopic ones like quantum dots \cite{740} and superconducting devices \cite{741}. In order to build a practical quantum computer, there are two broad criteria that must be fulfilled: 1) The coherence of superposition and entangled states must be maintained for periods of time long enough to complete a calculation without appreciable errors, and 2) the qubits must be individually controlled and manipulated within a large-scale architecture. Microscopic systems more readily fulfill the first criterion, as we routinely describe their behavior using quantum mechanics. However, it is extremely challenging to manipulate individual atomic-sized objects and to integrate them into an architecture of myriads of qubits. Macroscopic systems easily fulfill the second criterion since we can fabricate many qubits on a chip and address them with individual wires, but, concomitantly, their quantum mechanical behavior is easily destroyed through their stronger interactions with the environment, a process generically described as decoherence.

SMMs as qubits may offer the best of both worlds. With magnetic moments an order of magnitude larger than the moment of an electron, they may be easier to manipulate than atomic-sized objects. Yet their quantum behavior may mirror atomic-scale objects more than macroscopic ones.

SMMs have many important advantages as potential qubits. Many properties (e.g. barrier height, tunneling rate, interaction with environmental degrees of freedom) can be chemically engineered. Magnetic fields can be used to tune the barrier height and, in particular, the tunnel splittings. Moreover, microwave fields can be used to manipulate the quantum state of the spin and create superposition states.

Interest in using SMMs in quantum computing was galvanized by theoretical work by Leuenberger and Loss in 2001 \cite{298} that showed how Grover's quantum search algorithm could be implemented within a single, high-spin SMM using an elaborate superposition state (but no entanglement). Realizing that proposal is still well beyond current technology but activity in the field remains high. We now turn to a discussion of efforts to measure and exploit coherent quantum phenomena in SMMs.

\begin{figure}[tb]
\centering
\includegraphics[width=1\linewidth]{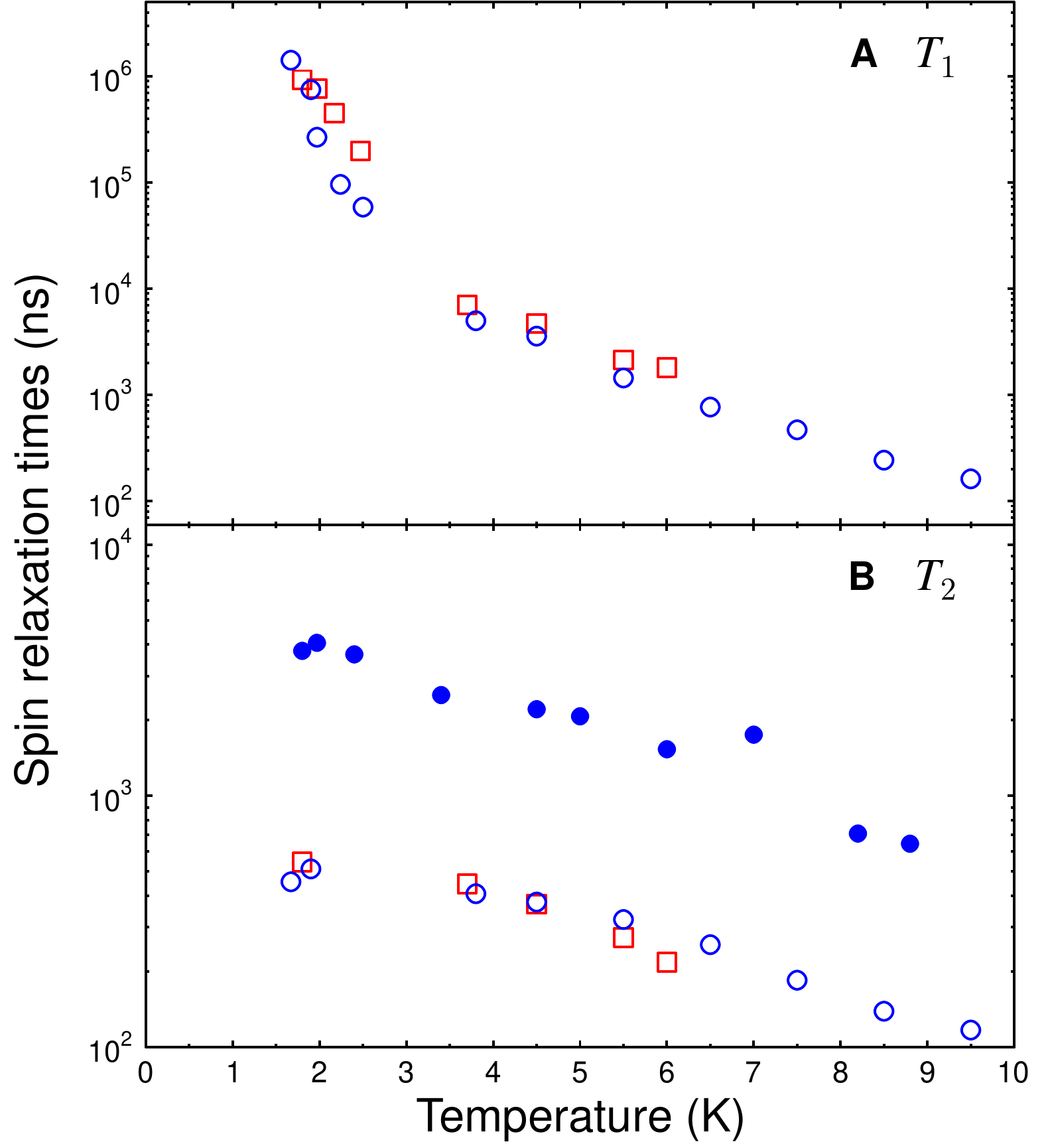}
\caption{Measured values of (a) the spin relaxation time $T_1$ and (b) the spin dephasing time $T_2$ as a function of temperature $T$ for Cr$_7$Ni (blue open circles) and Cr$_7$Mn (red squares). The blue filled circles show $T_2$ for a deuterated sample of Cr$_7$Ni. From Ref.~\cite{602}; reproduced with permission.}
\label{fig8}
\end{figure}

\begin{figure}[tb]
\centering
\includegraphics[width=1\linewidth]{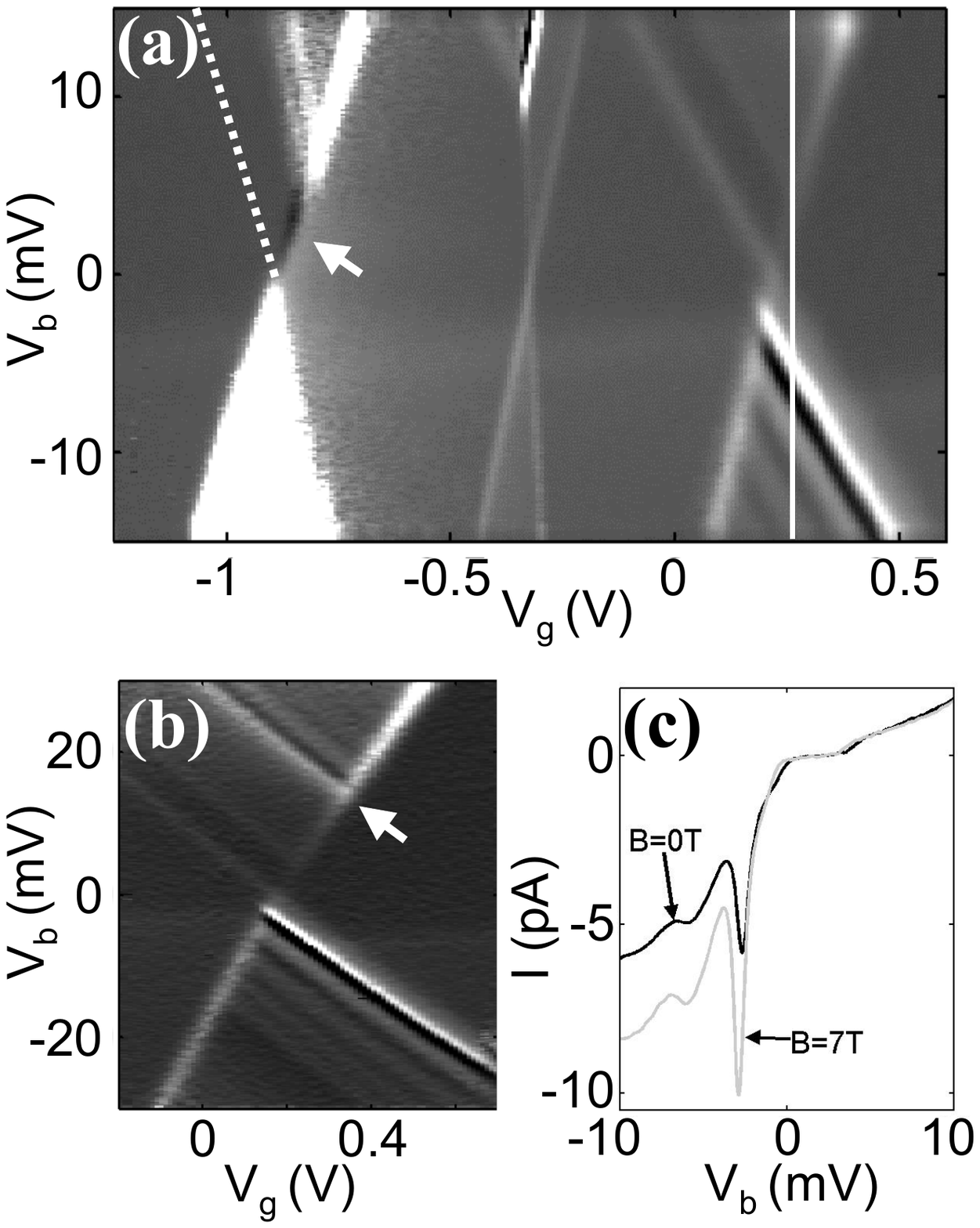}
\caption{(a) Grayscale plot of the differential conductance through an individual Mn$_{12}$-ac molecule as a function of bias voltage $V_b$ and gate voltage $V_g$. The arrow indicates a region of complete current suppression. (b) Same as (a) over a different range of $V_b$ and $V_g$. The arrow indicates a $14$ meV excitation characteristic of Mn$_{12}$ molecules used as an indicator of successful incorporation of the SMM into the transistor. (c) $I-V_b$ curves for a fixed value of $V_g$ (vertical white line in (a)) shows a region of negative differential conductance. From Ref.~\cite{773}; reproduced with permission.}
\label{fig9}
\end{figure}

Several groups have been studying the effects of microwave radiation on SMMs with the intent of observing coherent phenomena, such as Rabi oscillations or spin-echo effects. Early work showed photon-assisted tunneling effects and measured excited-state lifetimes \cite{283, 339, 337, 563, 585, 581, 657, 656, 700, 688}. A major advance was made in 2007 by Ardavan \emph{et al.}~\cite{602}, who observed spin-echo in Cr$_7$Ni and Cr$_7$Mn SMMs (as above, these names are shorthand for more complicated chemical formulas that, importantly, include many hydrogen ions). These are variants of Cr$_8$ wheels, which are antiferromagnetic with S=0. Substituting Ni or Mn for one Cr (as well as a compensating cation) results in a SMM with S=1/2 or 1, respectively. The former has no zero-field anisotropy ($\sigma_z^2=I$, the identity operator) and the orientation of the molecules with respect to the applied field is therefore irrelevant, except for a small anisotropy of the $g$ factor. This allowed Ardavan \emph{et al.}~to dilute the molecules in a solvent to the point that intermolecular dipole interactions were negligible. Using standard ESR techniques, they observed spin echos and measured excited-state lifetimes, $T_1$, as high as $\approx 1$ ms and dephasing times, $T_2$, of 100 - 1000 ns at low temperatures, as shown in Figure~\ref{fig8}. Interestingly, nearly identical results were obtained for the S=1 Cr$_7$Mn SMM (red open squares), despite the addition of anisotropy. A substantial component of the decoherence in these systems derives from hyperfine fields, as illustrated by the blue, filled circles in Figure~\ref{fig8}, which shows $T_2$ values for a deuterated sample of Cr$_7$Ni. The reduction in hyperfine fields by the substitution of deuterium for hydrogen raises the coherence time to $\approx$3 $\mu$s at low temperature. SMMs based on polyoxometalates have been suggested as qubit candidates because they lack nuclear spins \cite{743}.

Antiferromagnetic rings remain interesting low-spin systems and potential qubits. Recent experiments have shown that pairs of Cr$_7$Ni wheels can be exchange-coupled \cite{667} and there has been a theoretical proposal for turning a linear chain of rings into a scalable quantum processor \cite{744}.

Coherent phenomena have now been observed in a few other SMMs: V$_{15}$ \cite{695}, Fe$_4$ \cite{685} and Fe$_8$ \cite{672}. In the first two, this was achieved by diluting the molecules in solvent, a method similar to that used by Ardavan \emph{et al.} For Fe$_8$, Takahashi \emph{et al.}~\cite{672} used a large magnetic field and low temperatures to polarize the molecules within a single crystal so that each molecule experiences nearly the same local dipolar field. Also of note is recent work on Mn$^{2+}$ ions diluted in a MgO non-magnetic matrix \cite{674}. These single ions have a relatively large spin (S=5/2) and show coherent phenomena similar to SMMs.

The qubit state of an SMM need not be the spin state. SMMs that lack inversion symmetry \cite{682} can have definite chirality states. These states can be used as effective qubit states that can couple to electric fields through spin-orbit effects. Experimental evidence for chirality states was recently reported for a SMM triangle of three Dy ions \cite{692}.

Entanglement between qubits is essential for making a universal quantum computer. Entanglement may exist within the molecules, but if each SMM is to act as a single qubit, one is interested in entanglement \emph{between} SMMs. Passive entanglement (in which the energy eigenstates happen to be entangled states) has been observed in dimers of SMMs through magnetization \cite{443}, spectroscopy \cite{286} and specific heat measurements \cite{667}. As of this writing, controlled entanglement -- the creation of a well defined entangled state on demand through a controlled gate operation -- has not yet been achieved. There have been several theoretical proposals for creating entanglement between SMM qubits: using radiation pulses \cite{744}, by injection of a linker spin from an STM tip \cite{713}, or by coupling to rf electric fields in a cavity \cite{682}. While these methods have promise for testing fundamental physics and proof-of-principle demonstrations of entanglement, they are probably unworkable for a large-scale quantum computer.

\section{Addressing and manipulating individual molecules}

If SMMs are to be useful qubits, they need to be individually addressed and controlled. Other possible applications, such as using SMMs for magnetic memory or in spintronic devices, similarly require individual addressability. Efforts to achieve this are multidisciplinary, lying at the interface of the fields of supramolecular chemistry, molecular electronics, spintronics and quantum control. In this section, we briefly review experimental efforts to create single layers of SMMs on surfaces that can be individually addressed and to measure individual SMMs through transport techniques.

Since large anisotropy lies at the heart of most SMM behavior, it is essential that when depositing the molecules on a surface, this property be preserved and that the molecules have a well-controlled orientation. In addition, as evidenced by the effects of solvent disorder discussed above, it is also important that any symmetry breaking interactions be minimized. Given these constraints, it is not surprising that progress in depositing both chemically and magnetically intact SMMs on surfaces has been slow. Progress in this area has been outlined in recent review articles \cite{771, 706, 776}; we highlight a few salient points here and refer the reader to those reviews for more details.

Many efforts have focused on the deposition of Mn$_{12}$-ac and its variants. Techniques include vapor deposition and chemical self-assembly. While some of these techniques result in monolayers or thin films of molecules, there is evidence that in many cases the molecules are not all intact, e.g. small fragments are found by STM imaging and Mn$^{2+}$ ions are detected while the unperturbed compound contains only Mn$^{3+}$ and Mn$^{4+}$ ions. Moreover, many of the key magnetic characteristics, such as hysteresis and tunneling, are absent in these monolayers. Some progress has recently been made by Voss \emph{et al.}~who reported deposition of largely intact Mn$_{12}$ molecules using a ligand exchange process in which both the Au surface and the molecules are prefunctionalized \cite{772}. Using STM, they found a bandgap for the molecules consistent with that predicted by \emph{ab initio} calculations \cite{755}.

There has been progress in attaching other SMMs to surfaces, such as Cr$_7$Ni rings \cite{770} and Fe$_4$ tetrahedrons \cite{748a, 733}. The latter has proved to be very promising: it is the only SMM to remain magnetically intact, showing both hysteresis and evidence of tunneling after grafting to the surface, as indicated by X-ray magnetic circular dichroism and muon spin-rotation techniques. These techniques indicate that Fe$_4$ molecules on surfaces exhibit slow relaxation at low temperatures, although with a lower barrier than that found in bulk crystals \cite{748a, 733}. In another notable recent result, the Fe$_6$-POM molecule has recently been bonded to the surface of single-wall carbon nanotubes and magnetic measurements indicate that the molecules exhibit hysteresis and magnetization tunneling \cite{769}.

There have been a few attempts to incorporate an individual SMM in a transistor-like device in which the molecule is attached to two leads and the current-voltage characteristic is measured \cite{773, 774, 751}. An example is shown in Figure~\ref{fig9}, which shows the differential conductance of a Mn$_{12}$ molecule attached to two Au leads as a function of bias voltage and an external gate voltage capacitively coupled to the molecule \cite{773}. Complex behavior is found, including regions of negative differential conductance (see Figure~\ref{fig9}(b)). The results have been interpreted in terms of the high-spin states of Mn$_{12}$ with the interaction of electrons added to the molecule during conduction. It should be noted that neither in this nor similar studies by other groups has hysteresis been found in the transport properties as a function of applied magnetic field. This may be due to damage to the molecules similar to that found when they are deposited on surfaces. It may also be the result of strong coupling to the leads that distorts the molecule or otherwise breaks its symmetry.

The experimental results on transport through individual SMMs have stimulated a great deal of theoretical work \cite{767, 766, 759, 761, 763, 760, 765, 777, 756, 758, 752, 753, 752a, 749, 768}. These include proposals for observing a novel form of the Kondo effect \cite{766,777,756,758}, a geometric-phase modulation of transport \cite{761, 777, 756, 768}, spin filtering \cite{752} by the SMM, and current-induced switching of the SMM spin \cite{760, 753}.

Observing any of these effects may be challenging in light of the difficulty in making electrical contacts to SMMs without significantly perturbing them. However, recent results are encouraging \cite{748a, 733, 769} and some groups are working on other, less invasive techniques for measuring the properties of individual SMMs \cite{701}.

\section{Summary}

From an unheralded beginning in 1980, when the first molecular magnet was synthesized, activity in the field of SMMs has grown rapidly and now involves an unusually broad range of disciplines including physics, chemistry, material science, nanoscience and nanotechnology, spintronics, and quantum information. As new techniques (such as spin polarized STM, scanning magnetometry, magnetic circular dichroism) become available, they are being brought to bear to probe fundamental questions and to investigate the potential of SMMs for various applications. Work is proceeding along many different fronts. Chemists have expended a great deal of effort in a quest to find SMMs with larger anisotropy barriers to enable operation at higher temperatures; a new barrier-height record of $86.4$ K has recently been set in a $S=12$ [Mn$_6$] complex \cite{754}, toppling the record held by Mn$_{12}-ac$ and its variants. Much progress has been made depositing molecules on surfaces and measuring transport through individual SMMs with the goal of using them for both quantum and classical information processing and storage. A particularly attractive feature of SMMs is that they are amenable to engineering by chemical and self-assembly techniques to control and design desirable properties of the molecules and the interactions between them. These characteristics include spin, barrier height, hyperfine fields, dipole and exchange interactions and spin-phonon coupling.

In the space allotted to this review, it is impossible to cover all the interesting aspects of the field. For further information, the reader is referred to references \cite{friedmancollection,christoureview,gatteschi,sessolibook}.

\section{Acknowledgments}

JRF acknowledges the support of the National Science Foundation under grant no. DMR-0449516; MPS acknowledges the support of the National Science Foundation under grant DMR-0451605

\end{document}